# Reduction of the lasing threshold in optically pumped AlGaN/GaN lasers with two-step etched facets

Sergi Cuesta,[1, *] Lou Denaix,[1] Florian Castioni,[2] Le Si Dang,[3] and Eva Monroy[1]

[1] Univ. Grenoble-Alpes, CEA, Grenoble INP, IRIG, PHELIQS, 17 av. des Martyrs, 38000 Grenoble, France
[2] Univ. Grenoble-Alpes, CEA, LETI, 17 av. des Martyrs, 38000 Grenoble, France
[3] Univ. Grenoble-Alpes, CNRS, Institut Néel, 25 av. des Martyrs, 38000 Grenoble, France

Email: sergi.cuestaarcos@cea.fr

OrcID:
Sergi Cuesta: 0000-0003-0262-5875
Eva Monroy: 0000-0001-5481-3267

## Abstract

We report a two-step process to obtain smooth and vertical {10-10} *m*-plane facets in AlGaN/GaN separate confinement heterostructures designed to fabricate UV lasers emitting at 355 nm. The process consists in a dry etching by RIE-ICP combined with a crystallographic-selective wet etching process using a KOH-based solution. The anisotropy in the wet etching rates between the different crystallographic planes of the AlGaN structure, allows the fabrication of flat and parallel facets without a degradation of the multilayered ensemble. The optical performance of the lasers display a major improved when using the two-step process for the definition of the cavity, in comparison to cavities fabricated by mechanical cleaving, with the lasing threshold under optical pumping being reduced to almost half.

# 1. Introduction

Ultraviolet (UV) lasers find applications in fields like medical diagnosis and treatment (particularly in the domains of oncology and dermatology), laser lithography, micromachining or three-dimensional printing [1–3]. In this spectral region, the use of excimer lasers is widespread, despite their many restrictions due to the use of corrosive halogen gases. An alternative is frequency conversion based lasers, such as Nd:YAG, although they are limited in terms of flexibility in emission wavelength. To address these issues, there is a strong demand for compact semiconductor UV lasers, and AlGaN has proved to be a promising material for the fabrication of these devices due to its direct ultra-wide-bandgap, wavelength tunability, and feasibility of doping [4–7]. However, the fabrication of AlGaN laser diodes still presents important material challenges. The lattice mismatch in the (Al,Ga)N system result in structural defects (dislocations or cracks) generated by the need to release the accumulated elastic energy [8–11]. Furthermore, the increase of the dopant activation energy with the Al content results in high resistivity layers, which hampers carrier injection [4,12–16]. Alternative approaches using optical [17–21] or electron-beam pumping [22,23] are also under consideration.

In order to lower the lasing threshold, it is crucial to reduce the optical losses. In the case of edge emitting lasers, the fabrication of very smooth and vertical mirror facets has a huge impact in the laser performance [24]. Crystallographic mechanical cleaving is a common method for the fabrication of lasers [25–27], although it is not practical for mass production and it limits the range of possible cavity lengths. Furthermore, the strain fields due to the lattice mismatch in GaN/AlGaN separate confinement heterostructures (SCH) lead to non-ideal cleaved facets, due to the surface roughness induced by stress relaxation during the cleaving process. Also, mechanical cleaving is difficult to apply when the epilayers are grown heteroepitaxially on substrates whose cleaving planes are not aligned with those of the active layers, e.g. GaN or AlN on *c*-sapphire.

An alternative method for the definition of the cavity is the use of lithography and etching [28,29], which grants a precise geometry control, independent of the substrate. However, reactive ion etching (RIE) does not provide the vertical walls required to attain lasing, and a complementary step consisting in a crystallographic selective etching process is required to reduce mirror losses. In this sense, the exposure of AlGaN to TMAH, NaOH or KOH is known to lead to crystallographic selective etching rates [30–34]. Such etching processes have been described also as a method to obtain top-down nanowires with crystallographic smooth {10-10} *m*-plane sidewalls for photonic applications [35–37], as well as fin-shaped or nanowire-based field effect transistors [38–40].

It is often reported that the fabrication UV laser cavities includes a wet etching step, most often with TMAH [21,22,24,41,42] or KOH [29]. However, the question remains if this process can lead to comparable or improved mirror facets with respect to mechanical cleaving, or to what point it might limit the device performance. In this work, we report an optimized two-step etching process, involving inductively coupled plasma (ICP) RIE followed by wet etching in AZ400K (KOH containing developer), to fabricate smooth, vertical facets in AlGaN/GaN SCH designed to operate as electron beam pumped lasers. We compare the laser performance of the etched structures with that of mechanically cleaved cavities from the same wafer, to assess the improvement of reflectivity obtained with the etching process. We show that the lasing threshold almost halved by using the two-step etching process. This improvement is due to the fact that wet etching reveals chemically an atomically smooth *m*-plane facet, whereas the stress associated with mechanical cleaving, combined with the strong strain fields in these heterostructures, results in a partial relief of elastic energy by creation of nanofaceted surfaces.

## 2. Sample structure

This study focuses on two GaN/AlGaN laser structures which were grown by plasma-assisted molecular beam epitaxy on free-standing GaN substrates. In both architectures, described in figures 1(a) and (b), the active region consists of a 10-period GaN/Al$_{0.1}$Ga$_{0.9}$N multi-quantum well (MQW) inserted in an Al$_{0.1}$Ga$_{0.9}$N/Al$_{0.2}$Ga$_{0.8}$N waveguide. The layer thicknesses indicated in the figures were confirmed by XRD measurements. The design labeled "SCH" is a separate confinement heterostructure (SCH) with chemically sharp heterointerfaces, whereas "GRINSCH" refers to a sample in which the Al$_{0.1}$Ga$_{0.9}$N/Al$_{0.2}$Ga$_{0.8}$N interfaces are graded along 35.2 nm in order to implement a graded-index separate confinement heterostructure (GRINSCH). Such graded layers pursued an enhanced diffusion of carriers generated in the top cladding layers towards the MQW. A more detailed explanation of the growth conditions and the benefits of the GRINSCH can be found in our previous work [43], which contains simulations of the band diagram of both samples.

To visualize and validate the sample structure, the samples were analyzed using scanning transmission electron microscopy (STEM) in a probe-corrected TFS Titan Themis microscope operating at 200 kV. With this purpose, lamella specimens were prepared using a Ga$^+$ beam in a Zeiss Crossbeam 550 focused ion beam (FIB) scanning electron microscope (SEM). The voltage was progressively decreased from 30 kV to 2 kV in order to reduce beam damage and to obtain a sample thickness of about 70-90 nm. The panels on the left side of figures 2(a) and (b) present a cross-section image of a lamella extracted from the SCH and the GRINSCH samples, respectively, observed using high-angle annular dark field (HAADF) detector. The contrast of the image is due to the chemical composition, with dark contrast being observed in areas that are more Al rich (lower atomic number) and bright contrast in areas that are Ga rich (higher atomic number). At this scale, the MQW is hardly discernible due to the small QW width in comparison to the barriers, so that there is not enough chemical contrast with the inner cladding layers. In order to increase the signal-to-noise ratio, atomic-scale STEM-HAADF images were obtained by acquiring and aligning a stack of 40 frames of the area of interest, with a pixel size of approximately 23 pm and a pixel time of 200 ns. The zoomed images presented on the right side of figures 2(a) and (b) clearly reveal the contrast between the GaN quantum wells and the $Al_{0.1}Ga_{0.9}N$ barriers.

## 3. Mirror fabrication

In order to define the optical cavity of the laser structure, we need high reflectivity facets. Here, the facets were implemented by ICP etching followed by crystallographic selective wet etching. The dry etching step was performed in an ICP-RIE system using a $Cl_2$/$BCl_3$ (10/25 sccm) chemistry optimized for GaN (etching rate = 215 nm/min). The radiofrequency (RF) power was 220 W, the ICP power was 990 W, the sample temperature was 50°C and the pressure in the chamber was 10 mTorr. In a first approach, positive photoresist (AZ-5214E) was used as a mask for the definition of the cavities. In that case, an etching depth of 1 μm was easily obtained, but the etched sidewalls form an angle of 50° with the direction normal to the surface. To reduce this tapering, a second approach consisted in the implementation of a $SiO_2$/Ni (500 nm/150 nm) hard mask. For this purpose, a $SiO_2$ layer was deposited by chemical vapor deposition. Then, AZ-5214E negative photoresist was used to define the cavities by photolithography, followed by electron-beam evaporation of Ni and lift off. Finally, the $SiO_2$ mask was patterned by ICP-RIE using a CF$_4$/CH$_2$F$_2$ chemistry (25/25 sccm, RF power = 125 W, ICP power = 600 W, pressure in the chamber = 5 mTorr). The scanning electron microscopy (SEM) image in figure 3(a) presents a tilted view of the GRINSCH structure after dry etching. The top dark layer corresponds to the SiO$_2$/Ni hard mask. With this mask, the facet angle was reduced to 15°. The mask was removed by washing in HF.

The as-dry-etched samples did not display laser emission under optical pumping, which is explained by the imperfection of the optical cavity due to the tilted facet [26,44]. As modelled by L. van Deurzen, *et al.* [24], short wavelength emitters based on cavity optics have a high sensitivity to boundary imperfections. Therefore, a crystallographic-selective wet etching process is required to further reduce the facet tilt. The wet chemical etching of AlGaN in KOH-based solutions is highly anisotropic [33], and can be used to reduce the tapering.

For the optimization of the wet etching, initial tests were performed on a 4-µm-thick GaN-on-sapphire template grown by metalorganic vapor-phase epitaxy (MOVPE). The layer was first dry-etched following the above-described procedure, with the result shown in figure 3(b-c) after removal of the hard mask. Then, the sample was dipped in a deionized water solution of AZ400K developer (KOH-based developer from AZ Electronic Materials USA Corp.), AZ400K:H$_2$O = 1:4, stabilized at 65°C. Figure 3(d-i) shows the resulting facets after 2 h, 6 h and 10 h of chemical treatment. The images reveal how the facets evolve under wet etching. In a first step, the areas that were damaged during the dry etching process are rapidly removed, and we observe the emergence of nanofacets at the sidewalls (figures 3(d-e)), mostly *c*-(0001) and *m*-{10-10} planes. Increasing the etching time, the vertical facets become larger and the presence of (0001) nanofacets is restricted to the part of the sidewall closer to substrate (figures 3(f-g)). For longer etching time (figures 3(h-i)), vertical facets are obtained. Figure 3(j) is a top-view of the resulting facets, showing the straight morphology of *m*-{10-10} planes and the roughness of *a*-{11-20} planes due to the presence of *m*-{10-10} nanofacets.

The shape that results from the etching process is explained by the Wulff-Jaccodine model [45], which can predict the 3D geometry over time [46–48]. This geometrical model, which can be considered the inverse of a growth model, idealizes the crystal facets as mathematical planes, neglecting real surface structures. It considers that facets with slow etch rate appear in concave geometries, leading to better quality etched planes than in convex areas. In KOH-based solutions, AlGaN *m*-{10-10}, *a*-{11-20} and *c*-(0001) planes have etching rates with orders of magnitude slower (nm/s) than those of semipolar and (000-1) planes (µm/s). Therefore, vertical facets are exposed, particularly those with slowest etching rate, i.e. *m*-{10-10} planes [49].

This experiment was useful to visualize the process leading to smooth *m* facets. However, for device fabrication, the process should be accelerated. Therefore, for the fabrication of the laser mirrors, we have used non-diluted AZ400K stabilized at 80°C, with the results illustrated in figure 4(a-b) (GRINSCH sample). Note that the facets are vertical and there is no significant differences in the etching rate for the various layers in the stack.

To assess the optical performance of the etched laser structures (cavity length $L$ = 0.3 mm), we also prepared a series of laser bars with cavity lengths between 0.5 and 1.5 mm by mechanical cleaving along *m*-{10-10} planes, for both SCH and GRINSCH. The profile of the cleaved bars were imaged by SEM (figure 4(c)) and AFM (figure 4(d)), showing that the mechanical cleaving leads to tilted faceting due to the stress relaxation of the AlGaN/GaN heterostructure. The root mean square (rms) roughness of the facet at the level of the heterostructure is in the order of 20 nm, while the rms roughness of the facet generated at the level of the GaN substrate is in the order of 0.4 nm. Lasing was demonstrated in these laser bars (labeled S1 and S3 in ref. [43]), but the enhanced roughness at the heterostructure can surely produce significant light scattering when operating at short wavelengths [24].

## 4. Optical characterization

Photoluminescence (PL) measurements were performed under pulsed excitation using an Nd:YAG laser ($\lambda$ = 266 nm, 0.5 ns pulse width and 8 kHz repetition rate). The laser beam was shaped into a 100 μm wide stripe using a cylindrical lens. In length, the variation of the laser intensity was <10% over 1 mm. The samples were pumped perpendicular to the top surface, with the laser stripe perpendicular to the cavity mirrors. The PL edge emission was collected through a Jobin Yvon HR460 monochromator and detected by a UV-enhanced charge-coupled device (CCD) camera. Lasing threshold measurements of samples SCH and GRINSCH, were performed for different cavity lengths (0.5 mm, 0.75 mm, 1.0 mm and 1.5 mm) of mechanically cleaved laser bars and for 0.3-mm-long etched mesa cavities. All the measurements were performed at room temperature with the samples mounted on a copper holder.

Figures 5(a) and (b) show the PL spectra for a 0.3-mm-long GRINSCH cavity with etched facets and a 1-mm-long cavity with mechanically cleaved facets, respectively. The narrow line at 355 nm emerging at high pumping densities is assigned to laser emission. Figure 5(c) presents the PL intensity as a function of the excitation power density, showing a superlinear dependence above the lasing threshold, which is 100 kW/cm$^2$ for the etched cavity of 0.3 mm length and 180 kW/cm$^2$ for the cleaved cavity of 1 mm length. To get a deeper insight on the improvement of the optical performance using etched facets, the same optical measurements were performed for different cavity lengths on both SCH and GRINSCH cavities. The results are summarized in figure 5(d), which describes the evolution of the threshold as a function of the cavity length. Note that the GRINSCH structure has systematically lower power density threshold due to the improved carrier transfer to the active MQW [50].

The power density threshold is the pumping power density required to compensate the optical losses in the structure, including the internal ($\alpha_i$) and mirror ($\alpha_m$) losses. Therefore, the dependence of the threshold ($P_{th}$) on the cavity length ($L$) can be expressed as:

$$P_{th} = k(\alpha_i + \alpha_m) = k\left(\alpha_i - \frac{1}{2L}\ln(R_1 R_2)\right) \quad (1)$$

where $k$ is a proportionality constant and $R_1$ and $R_2$ are the reflectivity of the two cavity mirrors. In our case, we can assume that both cleaved facets are identical ($R_1 = R_2 = R$). In the case of long cavities, the contribution of the mirror losses becomes negligible compared to the internal losses. On the contrary, the mirror losses become dominant in short cavities.

In the figure 5(d), the cleaved cavities follow the tendency described by equation (1) (dashed-lines in the graph). The fact that data follow the 1/L dependence indicates that the cavity losses are dominated by mirror losses. The etched cavities, with a length of 0.3 mm, are significantly shifted with respect to that trend, showing a much lower lasing threshold than cleaved cavities. As the internal losses are expected to be the same in both cleaved and etched cavities, the different lasing threshold is due to the improved reflectivity of the mirrors in the etched cavities. Assuming that the reflectivity of the etched facet is close to its ideal value, $R_i$ = 21%, defined by the refractive index contrast between GaN ($n_{GaN}$ = 2.6915) [51] and air at the emission wavelength ($\lambda$ = 355 nm), we can estimate the reflectivity of the mechanically cleaved mirrors to be $R_{cleaved} \approx 3\%$.

Therefore, we conclude that in these strongly strained heterostructures, cleavage causes surface roughness due to stress relaxation that translates in boundary imperfections for the optical cavity, and the implementation of a two-step etching process for the fabrication of the laser mirrors has proved to be beneficial for the laser performance.

## 4. Conclusions

We have developed a two-step process combining ICP-RIE and KOH-based crystallographic selective wet etching for obtaining vertical and smooth facets in AlGaN/GaN separate confinement heterostructures designed to implement UV lasers emitting at 355 nm. Lasing threshold measurements under optical pumping proved that the mirror losses play a dominant role in the device performance of mechanically cleaved cavities. Such losses can be significantly reduced by using the two-step etching process, which leads to a reduction of the lasing threshold to almost half.

**Acknowledgements**

This work is supported by the French National Research Agency via the UVLASE program (ANR-18-CE24-0014), and by the Auvergne-Rhône-Alpes region (PEAPLE grant).

**Data availability statement**

The data that support the findings of this study are available upon reasonable request from the authors.

**Figure Captions**

**Figure 1.** Schematic description of the AlGaN/GaN (a) SCH and (b) GRINSCH samples. Starting from the GaN substrate, the layers are labeled bottom outer cladding (BOC), bottom inner cladding (BIC), multi quantum well (MQW), top inner cladding (TIC) and top outer cladding (TOC).

**Figure 2.** HAADF-STEM image of the (a) SCH and (b) GRINSCH sample observed along the [11-20] zone axis (lamella specimen thickness 70-90 nm). A zoom of four quantum wells in the MQW is marked in red.

**Figure 3.** SEM images of the etched facets: (a) GRINSCH sample after dry etching, before removing the $SiO_2$/Ni mask, and (b-i) a 4-μm-thick GaN-on-sapphire template (b-c) after removing the $SiO_2$/Ni hard mask, and after (d-e) 2 h, (f-g) 6 h, and (h-i) 10 h of wet etching in AZ400K:H2O (1:4) at 65°C. The scale and crystallographic axis are the same in (a-i). (j) Top-view of the etched sample showing straight, vertical *m* facets and rough *a* walls showing *m* nanofacets.

**Figure 4.** (a-b) SEM image of the *m* facet of a GRINSCH laser cavity after the two-step etching process (dry etching + crystallographic selective wet etching). (c) SEM and (d) AFM images of the *m* facet of a GRINSCH laser cavity after mechanically cleaving.

**Figure 5.** PL spectra as a function of the pumping power density: (a) GRINSCH structure with etched facets and a cavity length $L$ = 0.3 mm, and (b) GRINSCH structure with mechanically cleaved facets and $L$ = 1 mm. (c) Variation of the PL intensity as a function of the pumping power density of the GRINSCH sample with etched facets ($L$ = 0.3 mm) and mechanically cleaved facets ($L$ = 1 mm). The dashed line shows the slope of a linear trend. (d) Power density threshold as a function of the cavity length for the SCH and GRINSCH structures, including mechanically cleaved (solid symbols) and etched (hollow symbols) samples. Dashed lines illustrate the trend given by equation 1.

**Figure 1**

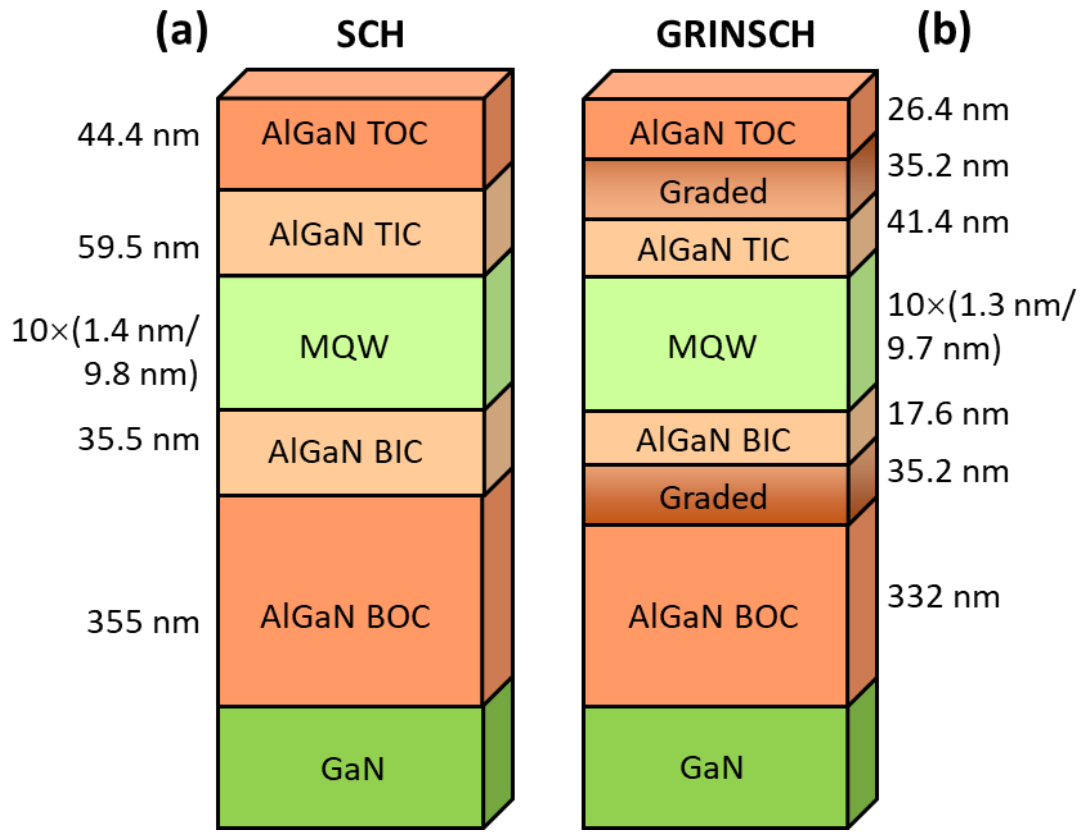

**Figure 2**

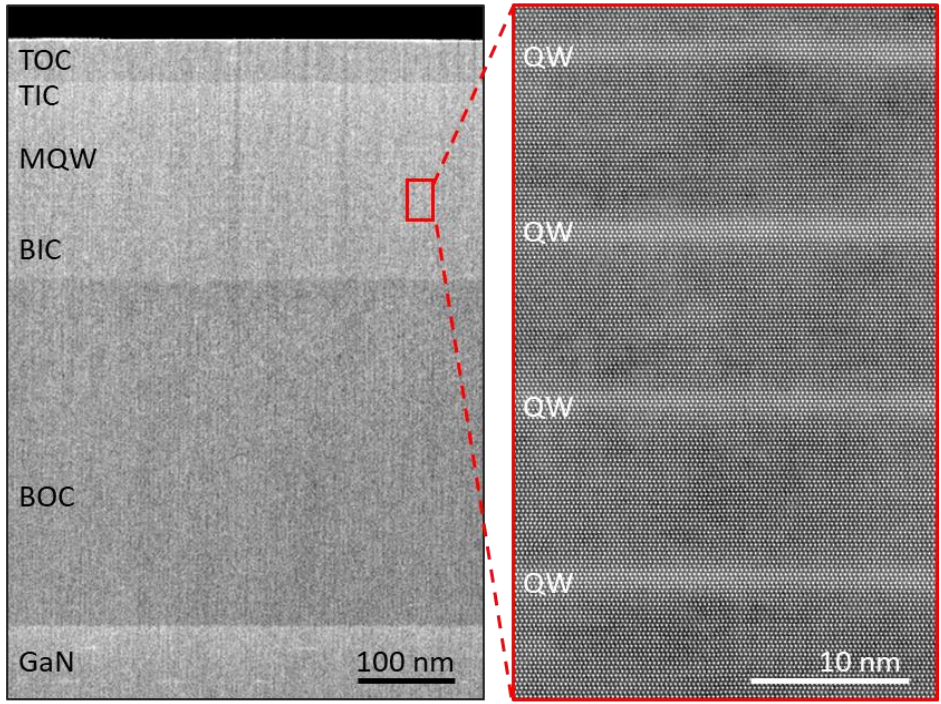

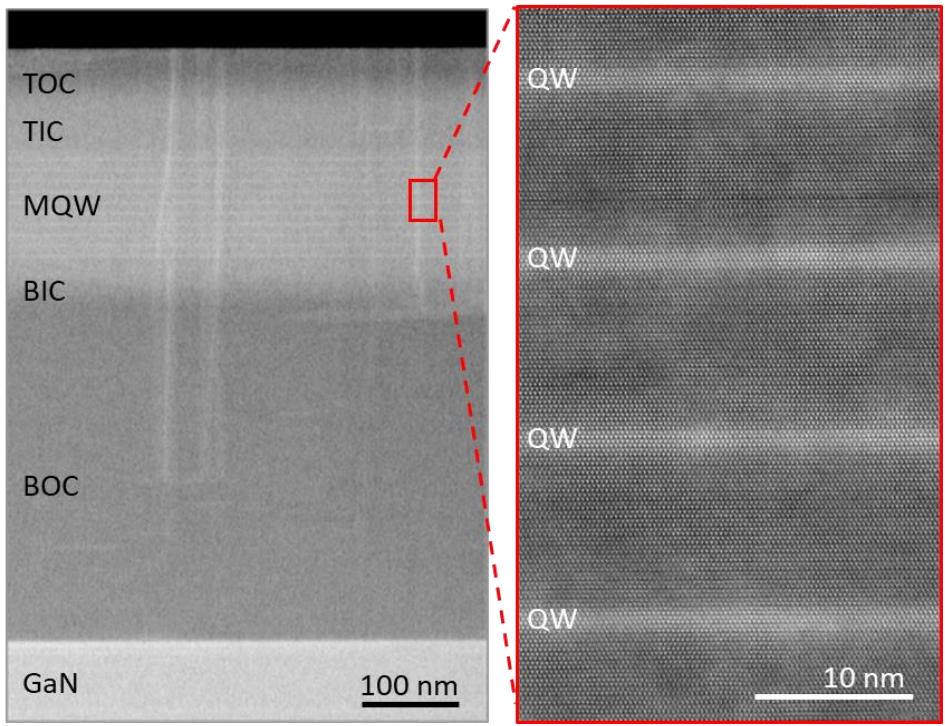

**Figure 3**

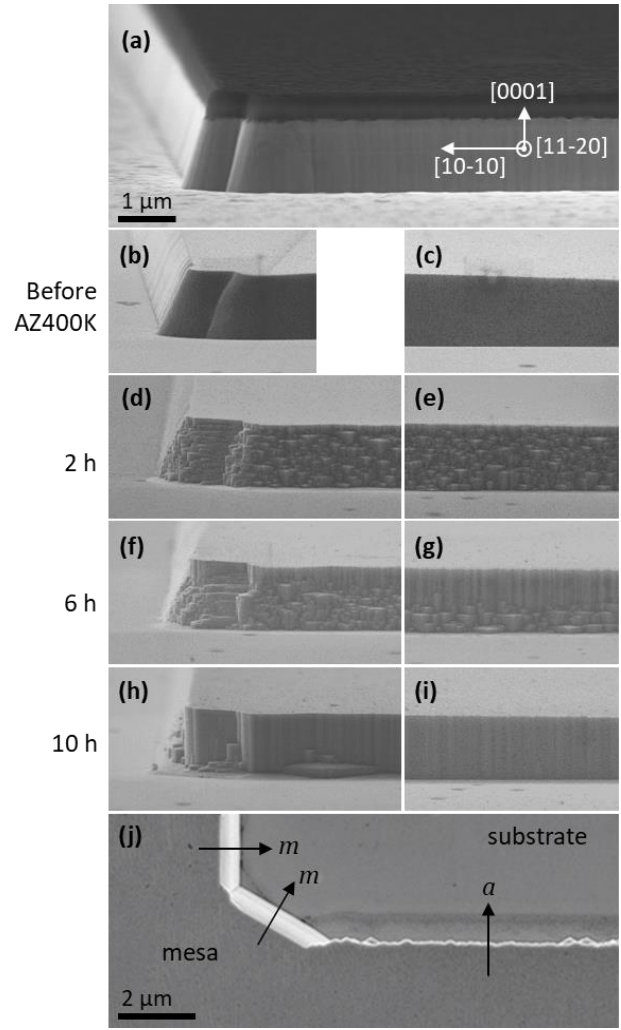

**Figure 4**

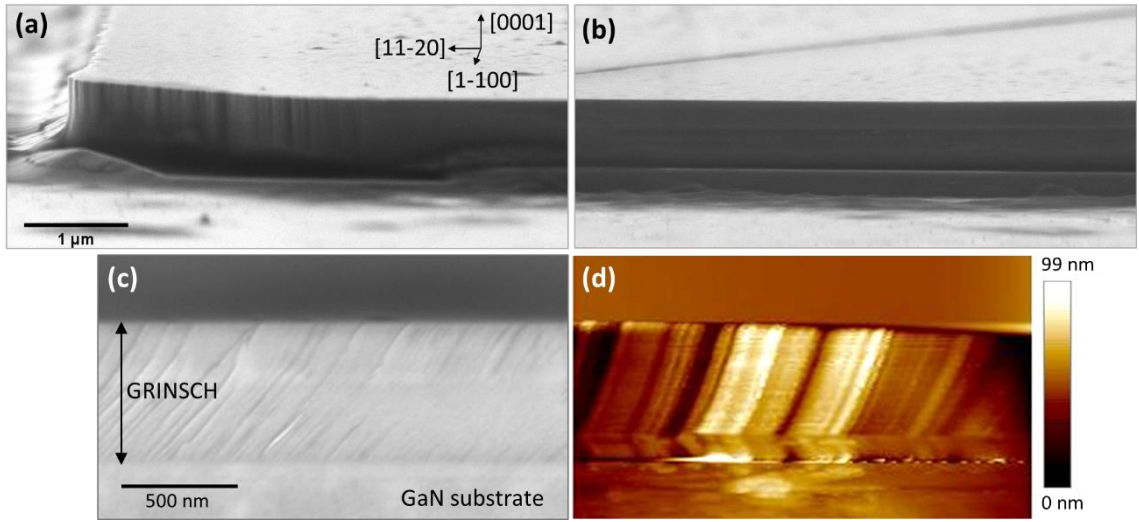

**Figure 5**

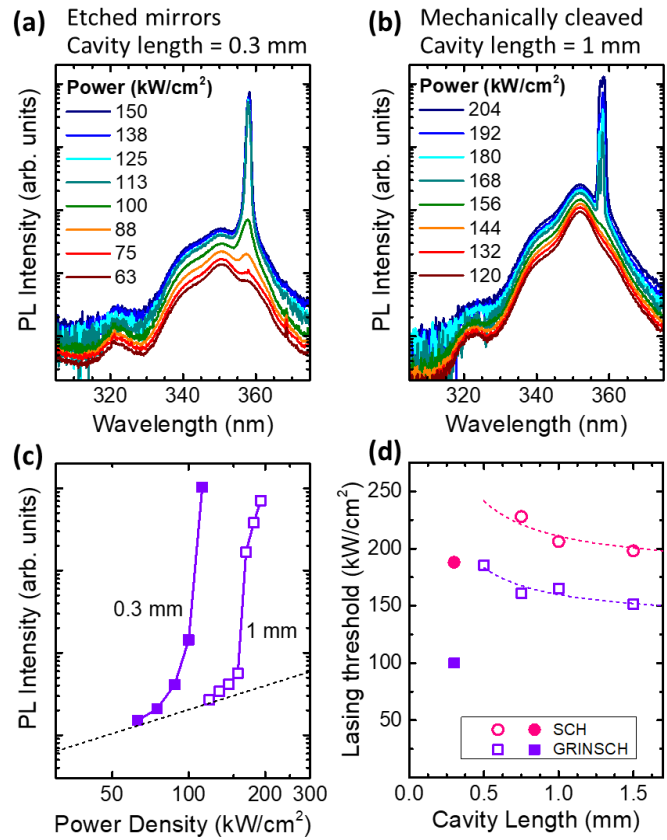